# Effects of initial compression stress on wave propagation in carbon nanotubes

M. M. Selim[1,2*], S. Abe[2] and K. Harigaya[2]

[1]*Department of Mathematics, AL-Aflaj community College, King Saud University, Al-Aflaj 710-11912 Saudi Arabia*

[2] *Nanotechnology Research Institute, AIST, Tsukuba 305-8568, Japan*

Abstract
An analytical method to investigate wave propagation in single- and double- walled carbon nanotubes under initial compression stress is presented. The nanotube structures are treated within the multilayer thin shell approximation with the elastic properties taken to be those of the graphene sheet. The governing equations are derived based on Flügge equations of motion. Frequency equations of wave propagation in single and double wall carbon nanotubes are described through the effects of initial compression stress and van der Waals force. To show the effects of Initial compression stress on the wave propagation in nanotubes, the symmetrical mode can be analyzed based on the present elastic continuum model. It is shown that the wave speed and amplitude ratio are sensitive to the compression stress especially for the lower frequencies.

1. Introduction
The deformation of carbon nanotubes (CNTs) has been the subject of numerous experimental, elastic continuum modeling, and molecular dynamics (MD) studies since the discovery CNTs [1-5]. Mechanical behavior of CNTs, including vibrational behavior, has been the subject of numerous recent studies [6-9]. Recently, some researches have been conducted on the vibration and wave propagation in CNTs to understand the dynamic behavior of CNTs [10-18]. The effect of initial stresses present in the medium is not considered in the above studies. The CNTs acting as basic elements of nano-structures often occur in initial stresses due to thermal stress, mismatch between different materials or initially external axial load. Some studies have taken the effect of initial stress on the wave propagation in CNTs [19-23].

Since CNTs are extremely small, the experiments to measure the properties of individual CNT are quite difficult. Therefore, the computational simulations have been regarded as a powerful tool to study properties of CNTs. There are two major categories for simulating the mechanical properties of CNTs: MD simulation and continuum mechanics. Although the MD method has been used for simulating the mechanical and physical properties of structures at atomic-scale level, this method is time-consuming and remains formidable especially for large-scale systems. Recently, solid mechanics with elastic continuum model have been regarded as an effective method and widely used for studying the mechanical and physical properties of CNTs [24-26]. For the effect of initial stress on the wave propagation of CNTs, prior studies presented the elastic wave solution obtained from the Bernoulli-Euler beam and the Timoshenko beam models. The effect of initial stress was limited to one-dimensional model since CNTs are modeled as beams. The Euler beam model ignored the influence of rotary and shear deformations on transverse wave propagation in CNTs. Wang and Varadan [12] have presented the elastic wave solution obtained from Euler– Bernoulli beam and Timoshenko beam models. They reported that the comparison between the two models could be inappropriate on the terahertz frequency range. This suggested that the Timoshenko beam model should be employed in analyzing the wave propagation for the high frequency range.

---

*Corresponding author; Fax: +96616826197; Email: msalim@ksu.edu.sa or selim23@ yahoo.com

Recently, the vibration of multi wall carbon nanotubes (MWCNTs) and wave propagation of double-wall carbon nanotubes (DWCNTs) have been studied based on Flügge shell equation [27-28]. These works show that the CNTs have the vibration and sound wave frequency over terahertz because of their nanoscale, which opens a new topic on wave characteristics.

The main aim of this work is to study effects of initial compression stress on wave propagation of single- and double-walled carbon nanotubes. The analysis is based on Flügge shell equation. As the wave numbers are the roots of an eighth order polynomial, Derive program version 6 was used to derive algebraic expressions for these roots and to simplify the Algebraic expressions via a series of substitutions. The expressions presented here are valid for circumferential modes $n = 0$. The results are shown graphically for different values of initial compression stress parameter.

## 2. Theory
*2.1 Equations of motion*

The cylindrical coordinates system used for describing the wave propagation in the tubes under initial compression stress is defined in Fig1. Approximate Flügge shell equations are proposed as the governing equations of the problem [29]. In the derivation of Flügge`s equations, the following assumptions for cylindrical shells are made:
1- All points that lie on a normal to a middle surface before deformation do the same after deformation.
2- Displacements are small compared to the shell thickness.
3- The normal stresses in the thickness direction are negligible (planar state of stress).
In particular, the first assumption may not correspond to reality in the neighborhood of the shell boundaries. However, this fact will not be considered here.

In the sequel, the $x$ coordinate is taken in the axial direction of the shell, where the $\varphi$ and $z$ coordinates are in the circumferential and radial directions, respectively (Fig.1). The displacements of the nanotube are defined by $u_1, u_2$ and $u_3$ in the direction of $x$, $\varphi$ and $z$ axes, respectively. The coordinates $u_1$ and $u_2$ represent in-plane axial and circumferential displacements of the tube wall midsurface, respectively, and $u_3$ represents the out-of-plane transverse displacement of the tube wall.

Based on the assumption stated above and using the coordinate system defined in Fig. 1, the Flügge type basic equations of motion are proposed as the governing equations of the wave propagation under initial compression stress ($P = -\sigma_{xx}$) in CNTs. The equations of motion for the nanotube can be written by

$$\left[\frac{\partial^2}{\partial x^2}+\frac{1-v}{2R^2}\frac{\partial^2}{\partial \varphi^2}-\beta\rho\, h\frac{\partial^2}{\partial t^2}\right]u_1+\left[\left(\frac{1+v}{2R}-\frac{\zeta_x E}{2R(1+v)}\right)\frac{\partial^2}{\partial x\partial\varphi}\right]u_2+\left[\left(\frac{v}{R}\frac{\partial}{\partial x}+\frac{\zeta_x E}{2R^2(1+v)}\right)\frac{\partial^2}{\partial\varphi^2}\right]u_3=0,$$

$$\left(\frac{1+v}{2R}\frac{\partial^2}{\partial x\partial\varphi}\right)u_1+\left[\left[\frac{1-v}{2}\frac{\partial^2}{\partial x^2}+\frac{1}{R^2}\frac{\partial^2}{\partial\varphi^2}+\sigma\left((1-v)\frac{\partial^2}{\partial x^2}+\frac{1}{R^2}\frac{\partial^2}{\partial\varphi^2}\right)-\beta\rho\, h\frac{\partial^2}{\partial t^2}\right]-\frac{\zeta_x E}{2(1+v)}\frac{\partial^2}{\partial x^2}\right]u_2$$

$$+\left[\frac{1}{R^2}\frac{\partial}{\partial\varphi}-\sigma\left((2-v)\frac{\partial^3}{\partial x^2\partial\varphi}+\frac{1}{R^2}\frac{\partial^3}{\partial\varphi^3}\right)+\frac{\zeta_x E}{2R(1+v)}\frac{\partial^2}{\partial x\partial\varphi}\right]u_3=0,$$

$$\left[\left(\frac{v}{R}\frac{\partial}{\partial x}\right)\right]u_1-\left[\frac{1}{R^2}\frac{\partial}{\partial\varphi}-\sigma\left((2-v)\frac{\partial^3}{\partial x^2\partial\varphi}+\frac{1}{R^2}\frac{\partial^3}{\partial\varphi^3}\right)+\frac{\zeta_x E}{2(1+v)}\frac{\partial^2}{\partial x^2}\right]u_2$$

$$-\left[\frac{1}{R^2}+\sigma\left(R^2\frac{\partial^4}{\partial x^4}+2\frac{\partial^4}{\partial x^2\partial\varphi^2}+\frac{1}{R^2}\frac{\partial^4}{\partial\varphi^4}\right)+\beta\rho\, h\frac{\partial^2}{\partial t^2}-\frac{\zeta_x E}{2(1+v)}\frac{\partial^2}{\partial x^2}\right]u_3+\hat{P}\beta=0,$$

(1)

where $h$ is the thickness of the CNT, $R$ is the radius of the middle surface of the tube, $\rho$ is the mass density, $E$ is the elastic modulus, $v$ is the Poisson's ratio, $\beta=\frac{(1-v^2)}{Eh}$, $\sigma=\frac{h^2}{12R^2}$ is the nondimensional thickness parameter, $\zeta_x=\frac{P(1+v)}{E}$ is the nondimensional initial stress parameter and $\hat{P}$ is the pressure between two adjacent nanotubes ( mainly due to the van der Waals interaction).

In matrix form, the Equations (1) may be expressed as:

$$\begin{bmatrix} D_{11} & D_{12} & D_{13} \\ D_{21} & D_{22} & D_{23} \\ D_{31} & D_{32} & D_{33} \end{bmatrix}\begin{bmatrix} u_1 \\ u_2 \\ u_3 \end{bmatrix}=\begin{bmatrix} 0 \\ 0 \\ -\hat{P}\beta \end{bmatrix},$$

(2)

where

$$D_{11}=\frac{\partial^2}{\partial x^2}+\frac{1-v}{2R^2}\frac{\partial^2}{\partial\varphi^2}-\beta\rho\, h\frac{\partial^2}{\partial t^2}, \qquad D_{12}=\left(\frac{1+v}{2R}-\frac{\zeta_x E}{2R(1+v)}\right)\frac{\partial^2}{\partial x\partial\varphi},$$

$$D_{13}=\frac{v}{R}\frac{\partial}{\partial x}+\frac{\zeta_x E}{2R^2(1+v)}\frac{\partial^2}{\partial\varphi^2}, \qquad D_{21}=\frac{1+v}{2R}\frac{\partial^2}{\partial x\partial\varphi},$$

$$D_{22} = \frac{1-\nu}{2}\frac{\partial^2}{\partial x^2} + \frac{1}{R^2}\frac{\partial^2}{\partial \varphi^2} + \sigma\left((1-\nu)\frac{\partial^2}{\partial x^2} + \frac{1}{R^2}\frac{\partial^2}{\partial \varphi^2}\right) - \beta\rho h\frac{\partial^2}{\partial t^2} - \frac{\zeta_x E}{2(1+\nu)}\frac{\partial^2}{\partial x^2},$$

$$D_{23} = \frac{1}{R^2}\frac{\partial}{\partial \varphi} - \sigma\left((2-\nu)\frac{\partial^3}{\partial x^2 \partial \varphi} + \frac{1}{R^2}\frac{\partial^3}{\partial \varphi^3}\right) + \frac{\zeta_x E}{2R(1+\nu)}\frac{\partial^2}{\partial x \partial \varphi}, \qquad D_{31} = -\frac{\nu}{R}\frac{\partial}{\partial x},$$

$$D_{32} = -\frac{1}{R^2}\frac{\partial}{\partial \varphi} + \sigma\left((2-\nu)\frac{\partial^3}{\partial x^2 \partial \varphi} + \frac{1}{R^2}\frac{\partial^3}{\partial \varphi^3}\right) + \frac{\zeta_x E}{2(1+\nu)}\frac{\partial^2}{\partial x^2},$$

$$D_{33} = -\frac{1}{R^2} - \sigma\left(R^2\frac{\partial^4}{\partial x^4} + 2\frac{\partial^4}{\partial x^2 \partial \varphi^2} + \frac{1}{R^2}\frac{\partial^4}{\partial \varphi^4}\right) - \beta\rho h\frac{\partial^2}{\partial t^2} - \frac{\zeta_x E}{2(1+\nu)}\frac{\partial^2}{\partial x^2}. \qquad (3)$$

Equation (2) represents a set of three expressions as:

$$\begin{aligned} D_{11}u_1 + D_{12}u_2 + D_{13}u_3 &= 0, \\ D_{21}u_1 + D_{22}u_2 + D_{23}u_3 &= 0, \\ D_{31}u_1 + D_{32}u_2 + D_{33}u_3 + \beta\hat{P} &= 0. \end{aligned} \qquad (4)$$

Eliminating $u_1$ and $u_2$ from the above equation, we get

$$D_1 u_3 + D_2(\hat{P}\beta) = 0. \qquad (5)$$

where

$$D_1 = D_{23}(D_{11}D_{32} - D_{12}D_{31}) + D_{13}(D_{31}D_{22} - D_{21}D_{32}) + D_{33}(D_{12}D_{21} - D_{11}D_{22}),$$

$$D_2 = (D_{21}D_{12} - D_{11}D_{22}). \qquad (6)$$

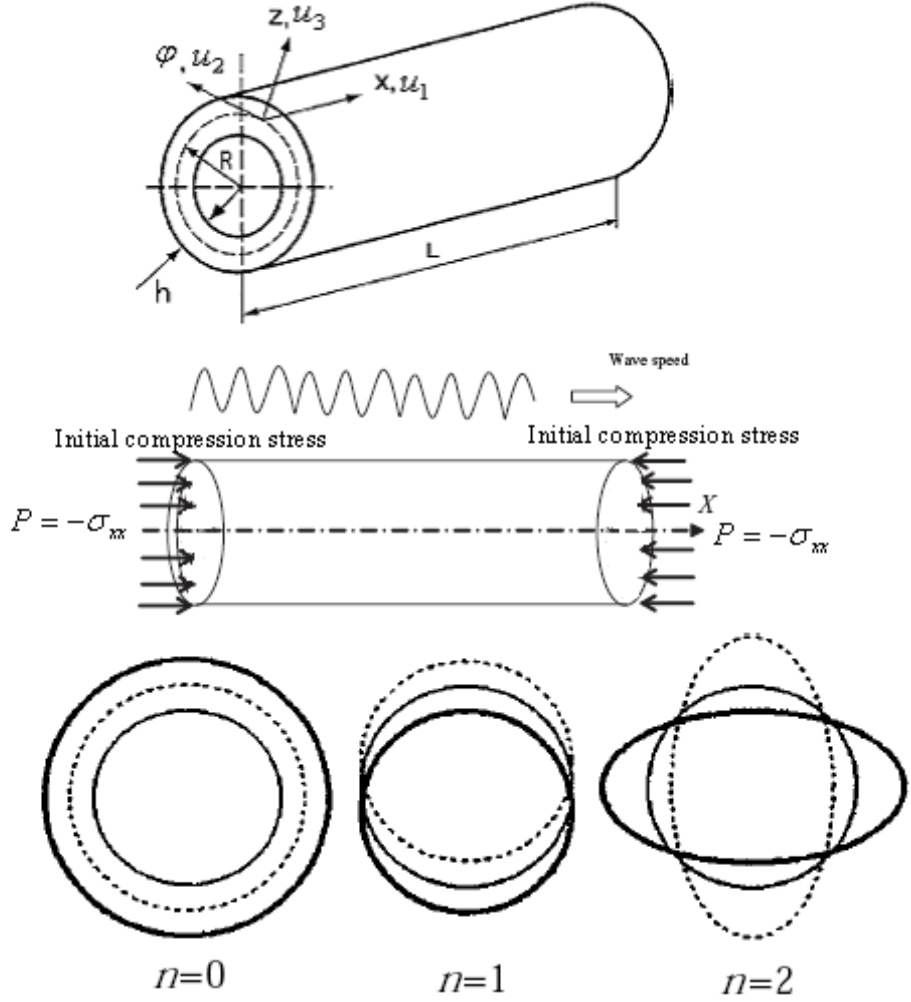

Fig.1. Geometry of the tube with coordinate system and modal shapes.

*2.2 Particular case*

When the effect of initial compression stresses is absent (i.e. $\zeta_x = 0$) equation (2) becomes

$$\begin{bmatrix} L_{11} & L_{12} & L_{13} \\ L_{12} & L_{22} & L_{23} \\ -L_{13} & -L_{23} & L_{33} \end{bmatrix} \begin{bmatrix} u_1 \\ u_2 \\ u_3 \end{bmatrix} = \begin{bmatrix} 0 \\ 0 \\ -\hat{P}\beta \end{bmatrix} \tag{7}$$

where

$$L_{11} = \frac{\partial^2}{\partial x^2} + \frac{1-v}{2R^2}\frac{\partial^2}{\partial \varphi^2} - \beta \rho h \frac{\partial^2}{\partial t^2}, \qquad L_{12} = \frac{1+v}{2R}\frac{\partial^2}{\partial x \partial \varphi}, \qquad L_{13} = \frac{v}{R}\frac{\partial}{\partial x}$$

$$L_{22} = \frac{1-v}{2}\frac{\partial^2}{\partial x^2} + \frac{1}{R^2}\frac{\partial^2}{\partial \varphi^2} + \sigma\left((1-v)\frac{\partial^2}{\partial x^2} + \frac{1}{R^2}\frac{\partial^2}{\partial \varphi^2}\right) - \beta \rho h \frac{\partial^2}{\partial t^2},$$

$$L_{23} = \frac{1}{R^2}\frac{\partial}{\partial \varphi} - \sigma\left((2-v)\frac{\partial^3}{\partial x^2 \partial \varphi} + \frac{1}{R^2}\frac{\partial^3}{\partial \varphi^3}\right),$$

$$L_{33} = -\frac{1}{R^2} - \sigma\left(R^2\frac{\partial^4}{\partial x^4} + 2\frac{\partial^4}{\partial x^2 \partial \varphi^2} + \frac{1}{R^2}\frac{\partial^4}{\partial \varphi^4}\right) - \beta\rho\, h\frac{\partial^2}{\partial t^2}, \qquad (8)$$

which agrees with the result of Natsuki et al.[14].

*2.3 Wave solution in single-walled nanotubes*
Nanotube wave propagation can be expressed in terms of two orthogonal wave components, one in the circumferential direction and the other in the axial direction. Only propagating waves with wave number $n$ occur in the circumferential direction resulting in circumferential mode shapes of the form $e^{in\varphi}$, where $\varphi$ is the nanotube angular position. In the axial direction, for a given circumferential mode $n$, the axial wave number is given by $k$, and the waves have the form $e^{ikx}$ where $x$ is the axial position along the nanotube axis. Then, the general solution of the wave propagation can be given by

$$u_3(x,\varphi,t) = A\exp[i(kx + n\varphi - \omega t)], \qquad (9)$$

where $A$ is the amplitude of the wave propagation, and $\omega$ is the circular frequency of the wave.

In case of single wall carbon nanotubes(SWCNT), the pressure $\hat{P}$ vanish (i.e. no interaction). In this case equation (5) takes the form
$$D_1 u_3 = 0. \qquad (10)$$
Thus, Substitution of Eq. (9) into Eq. (10) gives a polynomial function of $(\omega, k, n, \zeta_x)$. The characteristic equation
$$D_1(\omega, k, n, \zeta_x) = 0, \qquad (11)$$
represents the frequency equation of the phase velocity $(c_p = \omega/k)$ of the wave propagation in SWCNT under initial compression stress.

*2.4 Wave solution in double-walled nanotubes*
The DWNTs are assumed to be two individual coaxial tubes coupled together through van der Waals (vdW) interaction between the inner and outer nanotubes. The general solution form of the wave propagation can be given by

$$u_{3j}(x,\varphi,t) = A_j \exp[i(kx + n\varphi - \omega t)], \qquad (j=1,2), \qquad (12)$$

where $A_j$ (j = 1,2) are the amplitudes of the wave propagation in the inner and outer nanotubes. The amplitude ratio of the inner to the outer tubes is defined as $A_1/A_2$.

From the geometric structures, a DWCNT can be regarded as nested SWCNTs. The pressure ($\hat{P}$) between two adjacent nanotubes is mainly due to the van der Waals interaction. The vdW interaction energy potential, as a function of the interlayer spacing between the inner and outer nanotubes, can be estimated by the Lennard-Jones potential. In linear analysis, the vdW interaction pressure at any point between two adjacent tubes was modeled by a linear function of the deflection jump at that point [11]. In terms of the above model of the vdW interaction, the interaction pressure $\hat{P}$ in the Eq.(5) can be given by

$$\hat{P} = c(u_{32} - u_{31}), \tag{13}$$

where $c$ is the vdW interaction coefficient, approximately expressed as [30]

$$c = \frac{200 erg/cm^2}{0.16 d^2}, \quad (d = 1.42 \times 10^{-8} cm). \tag{14}$$

From Eq. (13) in Eq. (5), the coupled equations of the wave speed in double-walled nanotubes can written as

$$D_1^i u_{31} + D_2^i [\Delta(u_{32} - u_{31})] = 0,$$
$$D_1^o u_{32} + D_2^o [-\Delta(u_{32} - u_{31})] = 0, \tag{15}$$

where $D_1^i$ and $D_2^i$ are the differential operators for the inner tube, $D_1^o$ and $D_2^o$ are the differential operators for the outer tube defined in Eq(6) and $\Delta = c\beta$. Substituting Eq.(12) into Eq. (15), The phase velocity and the amplitude ratio can be determined by solving the two equations, which have only nontrivial solution at

$$\begin{vmatrix} D_1^i(\omega,k,n,\zeta_x) - \Delta D_2^i(\omega,k,n,\zeta_x) & \Delta D_2^i(\omega,k,n,\zeta_x) \\ \Delta D_2^o(\omega,k,n,\zeta_x) & D_1^o(\omega,k,n,\zeta_x) - \Delta D_2^o(\omega,k,n,\zeta_x) \end{vmatrix} = 0, \tag{16}$$

$$\frac{A_1}{A_2} = \frac{\Delta D_2^i(\omega,k,n,\zeta_x)}{\Delta D_2^i(\omega,k,n,\zeta_x) - D_1^i(\omega,k,n,\zeta_x)}.$$

It should be noted that the present model is only valid for small vibration of double-walled nanotubes due to the assumption of linearity of vdW force field. In terms of the eigenvectors given by Eq. (16), the modes of both nanotubes are coupled together; however, it is impossible to associate each of the modes to a specific nanotube even if the coupling may become small in some frequencies.

The characteristic equation

$$\begin{vmatrix} D_1^i(\omega,k,n,\zeta_x) - \Delta D_2^i(\omega,k,n,\zeta_x) & \Delta D_2^i(\omega,k,n,\zeta_x) \\ \Delta D_2^o(\omega,k,n,\zeta_x) & D_1^o(\omega,k,n,\zeta_x) - \Delta D_2^o(\omega,k,n,\zeta_x) \end{vmatrix} = 0 \tag{17}$$

represents the frequency equation of the phase velocity $(c_p = \omega/k)$ of the wave propagation in double- walled nanotube under initial compression stress.

3. Numerical results and discussion

The purpose of numerical computation is to analyze the effects of initial compression stress on wave propagation in CNTs. In order to perform numerical calculation the thickness of individual SWCNT was assumed to be that of a graphene sheet with 0.34 nm. CNTs had an elastic modulus of $(n=0)$ $E=1TPa\ 1$, and Poisson's ratio of $v = 0.27$, and the density $\rho = 3.34 g/cm^3$. The SWCNT was the $(40,0)$ zigzag tube with an effective diameter of 3.13 nm. The DWCNT consisted of the $(40,0)$ and $(50,0)$ zigzag tubes. The outer $(50,0)$ tube had an effective diameter of 3.91 nm. For the following solution we will just discuss the symmetric mode $(n=0)$ to show the effects of initial compression stress on the wave propagation in CNTs.

The dispersion curves of phase velocity of elastic waves with different frequencies for SWCNTs under initial compression stresses are presented in figures (2)a, 2(b) and 2(c). In Fig.2 (a), the variation of wave speed versus the frequency without initial stress $(\zeta_x = 0.0)$ is plotted. In order to examine the influence of the initial compression stresses on the wave propagation of SWCNTs, the results under initial compression stress $\zeta_x = 0.01$, $\zeta_x = 0.025$ are compared with corresponding ones without initial stress $(\zeta_x = 0.0)$ in figures. 2(a) and 2(b) respectively. Comparisons of figure 2(b) with figure 2(a) show that the change tendency of the wave speeds in single-walled CNTs with initial compression stresses $\zeta_x = 0.01$ is similar to those in single-walled CNTs without initial compression stresses $\zeta_x = 0.0$, but a change is observed in figure 2(b) for $\zeta_x = 0.025$ at frequency interval [6,8] THz.

Figures 3(a), 3(b) and 3(c) show the effect of initial compression stresses with different frequencies for DWCNTs. In Fig.3 (a), the variation of wave speed versus the frequency without initial stress $(\zeta_x = 0.0)$ is plotted. There are two different wave speeds given by equation (16). Comparisons of figures 3(b) and 3(c) with figure 3(a) show that the wave speeds in double-walled CNTs with initial compression stresses $\zeta_x = 0.01$ and $\zeta_x = 0.025$ are sensitive to the initial compression stress, especially for the frequencies interval [1,5]THz. It is also noted that, the peaks of the wave speed and its frequency are affected by the initial compression stress $(\zeta_x = 0.025)$, and the peak values are approximately 12 km/s with frequency of about 4 THz.

Figures 4(a), 4(b) and 4(c) show the amplitude ratios of the inner tube to the outer tube for the two wave speeds given by equation (16) for different initial compression stress values. Here, it is noticed that the effects of initial compressive stress on wave propagation in a double-walled CNTs effect the amplitude of wave speeds in the frequencies interval [1, 5] THz. The amplitude of wave speeds decreases as the initial compressive stress increases, and the corresponding amplitude ratios of the inner tube to the outer tube increase as the initial compressive stress increases.

4. Conclusion

This paper studies the effects of initial compression stress on wave propagation of single- and double-walled carbon nanotubes based on Flügge shell equation. It is shown that the wave speed and amplitude ratio are sensitive to the compression stress especially for the lower frequencies. The calculations were made for the symmetric mode and it is recommended to do these calculations for higher mode. The investigation presented may be helpful in the application of CNTs, such as high – frequency oscillators and mechanical sensors.

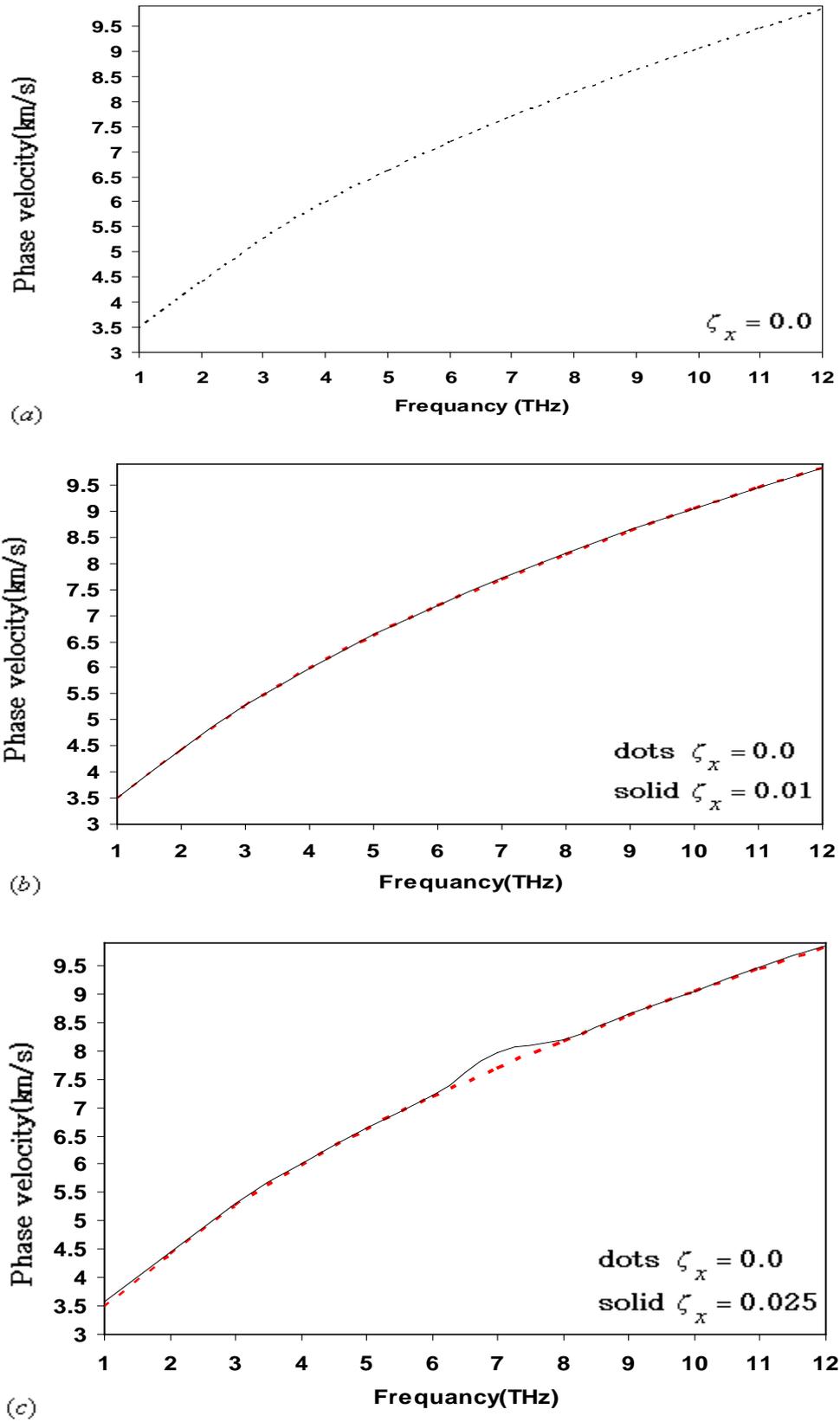

Fig.2. Dispersion curves of wave propagation for a single- walled carbon nanotube under initial compression stress: (a) initial stress parameter $\zeta_x = 0.0$, (b) initial stress parameter $\zeta_x = 0.01$ and (c) initial stress parameter $\zeta_x = 0.025$.

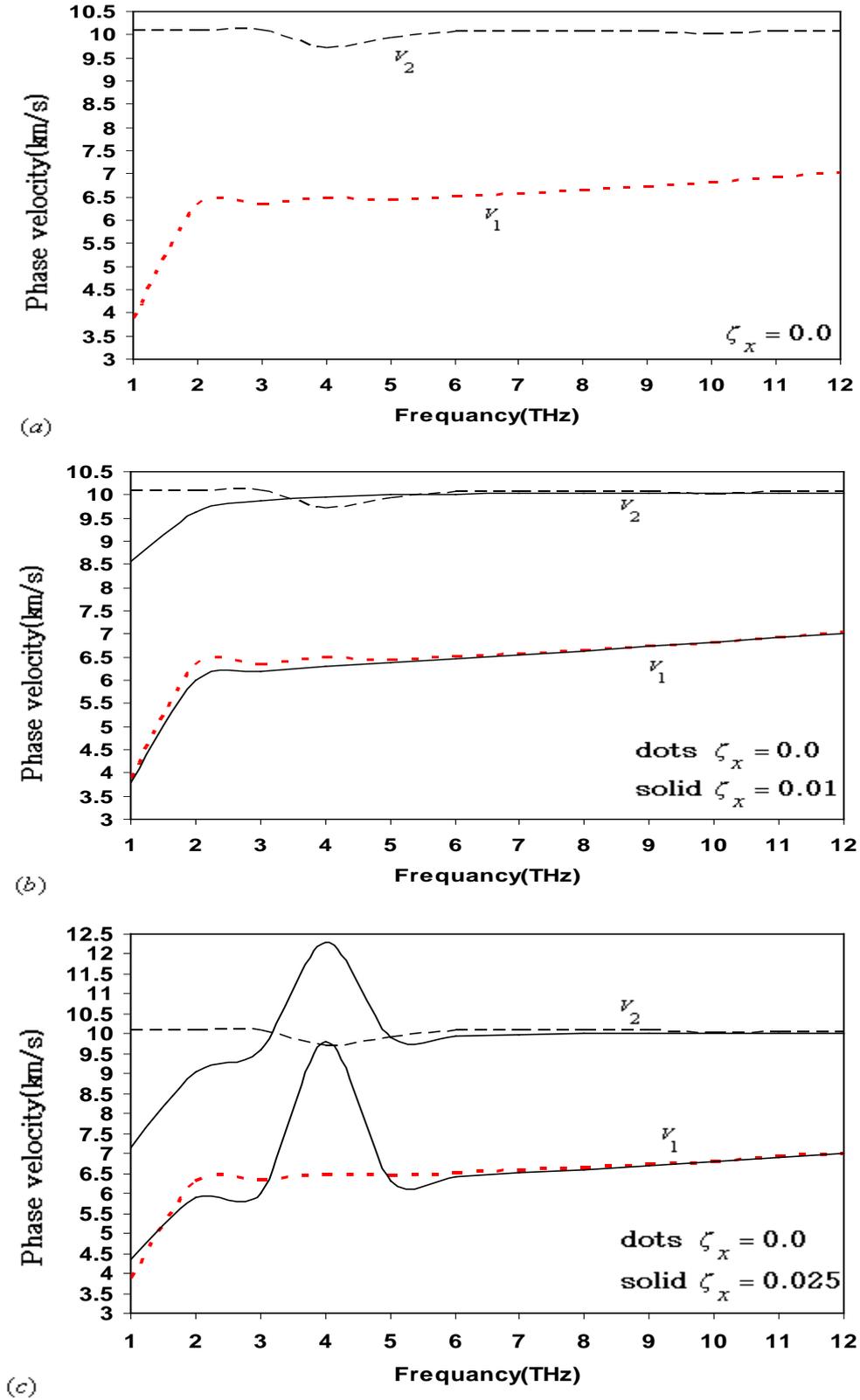

Fig.3. Dispersion curves of wave propagation for a double- walled carbon nanotube under initial compression stress: (a) initial stress parameter $\zeta_x = 0.0$, (b) initial stress parameter $\zeta_x = 0.01$ and (c) initial stress parameter $\zeta_x = 0.025$, where $V_1$ and $V_2$ express different wave speeds.

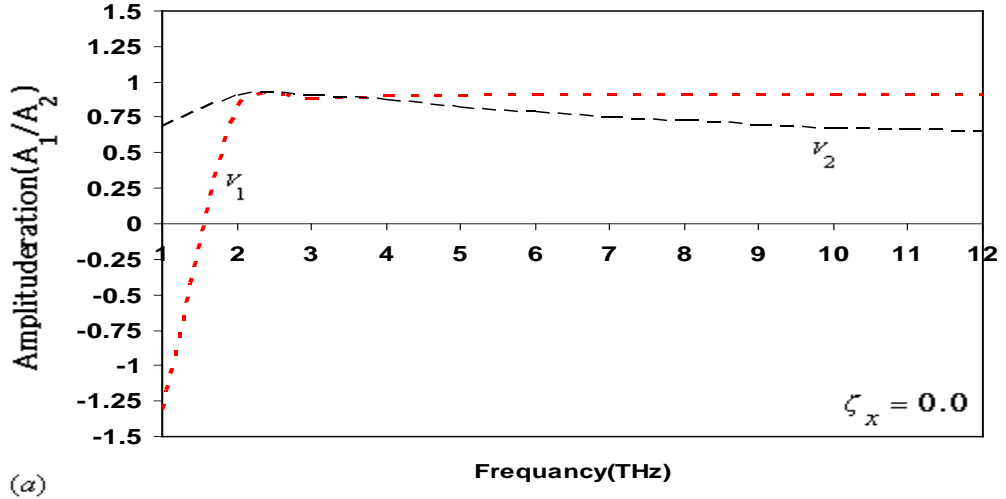

(a)

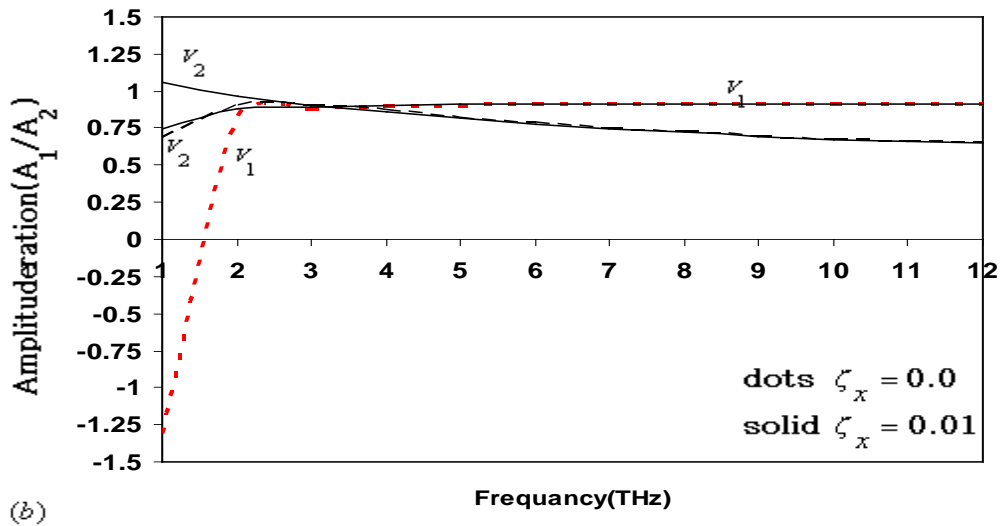

(b)

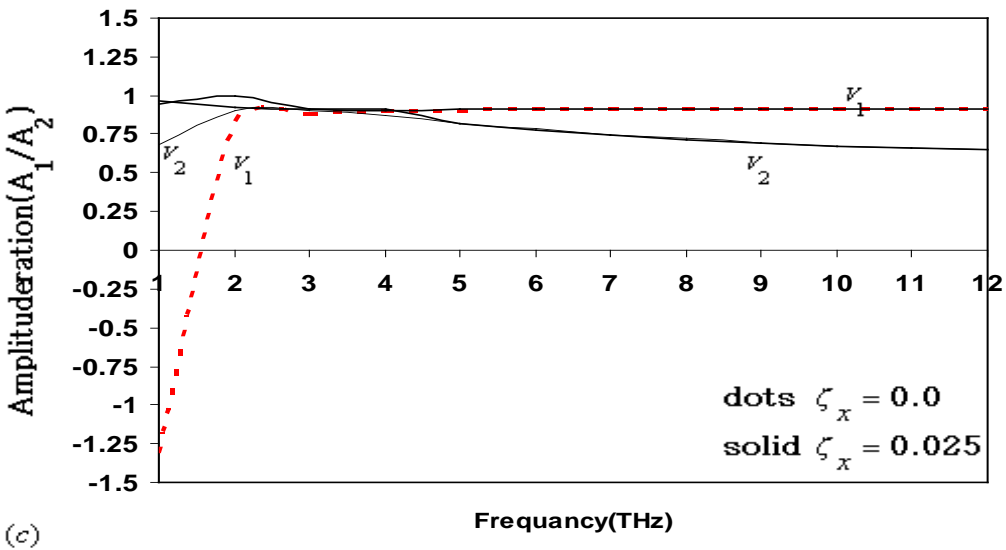

(c)

Fig.4. Amplitude ratio of phase velocity in a double- walled carbon nanotube under initial compression stress: (a) initial stress parameter $\zeta_x = 0.0$, (b) initial stress parameter $\zeta_x = 0.01$ and (c) initial stress parameter $\zeta_x = 0.025$, where $v_1$ and $v_2$ express different wave speeds.

ACKNOWLEDGMENT

This work was supported by the King Abdallah Institute for Nano Technology; King Saud University (Saudi Arabia).The first author would like to acknowledge the supporting and encouragement of Dr. Salman Alrokayan (Vice president of King Abdallah Institute for Nano Technology) during this work.